\documentclass[aps,prb,twocolumn,superscriptaddress,showpacs,floatfix]{revtex4}
\usepackage{graphicx}
\usepackage{latexsym}
\usepackage{stmaryrd}
\usepackage{amsmath}
\usepackage{amssymb}
\usepackage{epstopdf}
\usepackage[colorlinks,linkcolor=red,anchorcolor=red,citecolor=blue]{hyperref}

\begin{document}
\makeatletter
\newcommand{\rmnum}[1]{\romannumeral #1}
\newcommand{\Rmnum}[1]{\expandafter\@slowromancap\romannumeral #1@}
\makeatother

\title{Real space topological invariant and higher-order topological Anderson insulator in two-dimensional non-Hermitian systems}

\author{Hongfang Liu }
\affiliation{School of Physical Science and Technology, Soochow University, Suzhou, 215006, China}
\author{Ji-Kun Zhou }
\affiliation{School of Physical Science and Technology, Soochow University, Suzhou, 215006, China}
\author{Bing-Lan Wu }
\affiliation{School of Physical Science and Technology, Soochow University, Suzhou, 215006, China}
\author{Zhi-Qiang Zhang }\email{zqzhang2018@stu.suda.edu.cn}
\affiliation{School of Physical Science and Technology, Soochow University, Suzhou, 215006, China}
\author{Hua Jiang}\email{jianghuaphy@suda.edu.cn}
\affiliation{School of Physical Science and Technology, Soochow University, Suzhou, 215006, China}
\affiliation{Institute for Advanced Study, Soochow University, Suzhou 215006, China}
\date{\today}

\begin{abstract}
We study the characterization and realization of higher-order topological Anderson insulator (HOTAI) in non-Hermitian systems, where the non-Hermitian mechanism ensures extra symmetries as well as gain and loss disorder.
  We illuminate that the quadrupole moment $Q_{xy}$ can be used as the real space topological invariant of non-Hermitian higher-order topological insulator (HOTI). Based on the biorthogonal bases and non-Hermitian symmetries, we prove that $Q_{xy}$ can be quantized to $0$ or $0.5$. Considering the disorder effect, we find the disorder-induced phase transition from normal insulator to non-Hermitian HOTAI. Furthermore, we elucidate that the real space topological invariant $Q_{xy}$ is also applicable for systems with the non-Hermitian skin effect. Our work enlightens the study of the combination of disorder and non-Hermitian HOTI.
\end{abstract}

\maketitle

\section{Introduction}\label{S1}

With the development of topological theory of Hermitian systems, many kinds of topological states have been proposed and realized\cite{TI1,TI2,TI3,TI4,TI5,TI6,TI7,TI8,TI9,nature1,nature2,nature3,Fu1,Fu2}, among which the non-Hermitian counterparts\cite{NH1,NH2,NH3,NH4,NH5,NH6,NH7,NH8,NH9,NH10,NH11,NH12,NH13} have attracted great attention. Compared with the Hermitian cases, the non-Hermitian topological states are based on the non-Bloch theory\cite{NH6,NH7,NH8,NH9} due to the unique features of non-Hermitian Hamiltonian\cite{NH5,NH6,NH7,NH8,NH9,NH10,NHSE2,NHSE4,NHSE5,NHSE6}.
Soon after the establish of the non-Hermitian topological band theory, there is a blooming investigations of disorder effect\cite{NHSE6,dis1,dis2,dis3,dis4,dis5,dis6,NHD1,NHD2,IPR1,XunlongLuo} in non-Hermitian systems. By using the non-commutative geometry method\cite{dis5,dis6,NGM2,NGM3,NGM4,NGM5,NGM6,NHD1,NHD2}, the disorder-induced phase transitions of non-Hermitian Chern insulator\cite{dis5,dis6} and non-Hermitian Su-Schrieffer-Heeger model\cite{NHD1,NHD2} have been studied. The non-Hermitian topological Anderson insulator\cite{dis5,dis6,NHD1,NHD2} is also reported.

In these years, the Hermitian higher-order topological insulator (HOTI)\cite{HOTI1,HOTI2,HOTI3,HOTI4,HOTI5,HOTI6,HOTI7,HOTI8,HOTI9,Qxy1,Qxy2,Qxy3,Qxy4} is one of the most focus of topological states. Not only the nested-Wilson-loop method\cite{HOTI2,HOTI3,HOTI4} but also the real-space topological invariant\cite{HOTI9,Qxy1,Qxy2,Qxy3,Qxy4} were reported to characterize such states. Specially, in two-dimensional systems, the quadrupole moment $Q_{xy}$ \cite{HOTI9,Qxy1,Qxy2,Qxy3,Qxy4} was proposed to be the real space topological invariant of HOTI. Initially, the quantization of this topological invariant was thought to be protected by the point group in the HOTI\cite{HOTI4} which is considered as the topological crystal insulators\cite{Fu1}. However, Li {\it et al.} \cite{Qxy2}showed that the quantized $Q_{xy}$ could also be protected by chiral symmetry or particle-hole symmetry in Hermitian systems. Such results still hold even with disorder effects. They also carefully studied the disorder-induced phase transition in the corresponding systems and predicted the existence of a higher-order topological Anderson insulator (HOTAI).

Very recently, the interplay of HOTI and non-Hermitian is proposed and gains
extensive interests\cite{NH10,NH11,NHHO1,NHHO2}. The topological invariants in momentum space to distinguish such topological phases are also investigated\cite{NH10,NH11}. However, the researches referred to real space topological invariant and disorder-induced topological phase transition of non-Hermitian HOTI are seldom reported.
Since the Hamiltonian $H\neq H^\dagger$, the chiral symmetry or particle-hole symmetry in Hermitian cases will change into the distinctive symmetries\cite{NH13} in non-Hermitian cases. Moreover, disorders with gain and loss are also available for non-Hermitian samples. These features should affect the disorder effect and the related phase transitions, which is unique for non-Hermitian systems.

In this paper, we propose that the quadrupole moment $ Q_{xy}$, defined in the frame of the biorthogonal bases\cite{biorthogonal,NH6,NH7,dis6}, can be considered as the real space topological invariant of non-Hermitian HOTI. We prove that $Q_{xy}$ is quantized to $0$ or $0.5$ if there is a real diagonal matrix connecting Hamiltonian $H$ with $-H^\dagger$, which is universal for non-Hermitian samples with line gap along the real axis.
Specifically, taking the pseudoanti-Hermiticity symmetry \cite{NH10} as an example, we demonstrate that the quantized $ Q_{xy}=0.5$ can be utilized to identify the non-Hermitian HOTI. In terms of $ Q_{xy}$, the disorder-induced topological phase transition in non-Hermitian HOTI is also studied. With the help of the pseudoanti-Hermiticity protected quantization of $Q_{xy}$, we uncover that the non-Hermitian HOTI is robust against disorder. More importantly, the non-Hermitian HOTAI is predicted in such a system.
Finally, we manifest that $Q_{xy}$ is also applicable for systems with non-Hermitian skin effect (NHSE)\cite{NH5,NH7,NHSE1,NHSE2,NHSE4,NHSE5,NHSE6}.

This paper is organized as follows: In Sec. \ref{S2}, we present the method and the model. We demonstrate the realization of non-Hermitian HOTAI in Sec. \ref{S3}.
Then, a brief discussion and summary are presented in Sec. \ref{S4}.

\section{method and model}\label{S2}

\subsection{Quantized quadrupole moment protected by pseudoanti-Hermiticity symmetry}

\begin{figure}[t]
   \centering
    \includegraphics[width=0.42\textwidth]{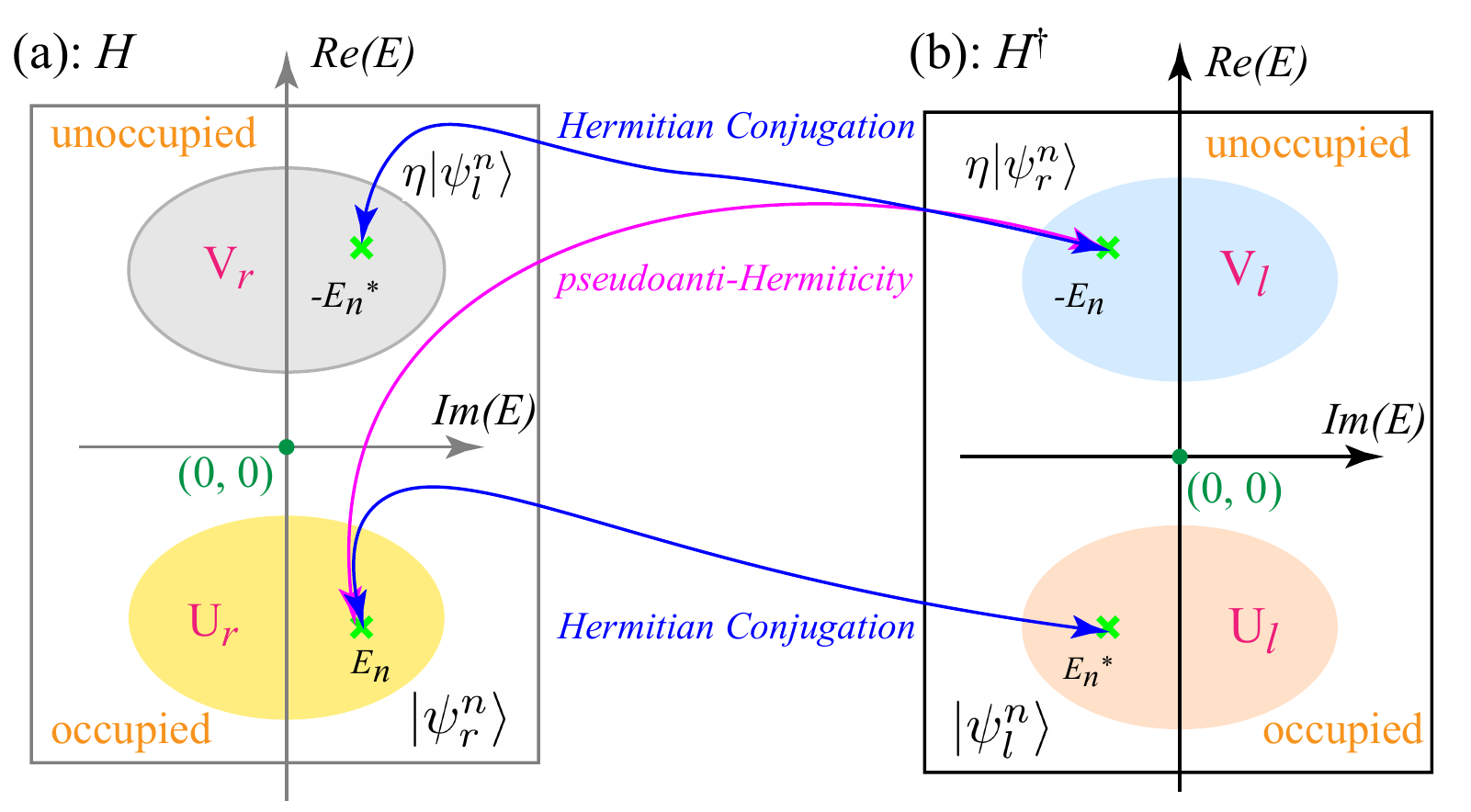}
    \caption{(Color online). The relationship between ${\it n_{th}}$ eigenvalues $E_n$, $E^*_n$, $-E^*_n$, and $-E_n$.
    The corresponding eigenvectors are $|\psi^n_r\rangle$, $|\psi^n_l\rangle$, $\eta|\psi^n_l\rangle$, and $\eta|\psi^n_r\rangle$.
    (a)$E_n$ and $-E^*_n$ are the eigenvalues of $H$. (b)$E^*_n$ and $-E_n$ are the eigenvalues of $H^\dagger$. $E_n$ ($-E_n$) and $E^*_n$ ($-E^*_n$) are connected by the Hermitian conjugation.
     The Hermitian conjugation corresponds to the transformation between Hamiltonian $H$ and $H^\dagger$, in which $|\psi^n_r\rangle$ will transform into $|\psi^n_l\rangle$.
     $E_n$ ($E^*_n$) and $-E_n$ ($-E^*_n$) are correlated by pseudoanti-Hermiticity.
      We define the occupied ($Re[E]<0$) and unoccupied ($Re[E]>0$) conditions with the help of line gap along axis $Re[E]$. The matrices constructed by the occupied (unoccupied) standard right and left eigenvectors are marked as $U_r$ ($V_r$) and $U_l$ ($V_l$), respectively.  }
   \label{f1}
\end{figure}

We suppose a non-Hermitian Hamiltonian $H$ holds the pseudoanti-Hermiticity symmetry $\eta$ with $\eta=\eta^{-1}$ and $\eta H\eta =-H^\dagger$. Such symmetry extends the chiral symmetry in Hermitian cases\cite{NH13}.
 According to the properties of Hamiltonian in non-Hermitian systems, the eigenvalues of $H$ and $H^\dagger$ are given as follows\cite{biorthogonal}:
\begin{equation}
H|\psi^n_r\rangle=E_n|\psi^n_r\rangle,  ~H^\dagger|\psi^n_l\rangle=E_n^*|\psi^n_l\rangle,
\label{EQ0}
\end{equation}
where $|\psi^n_r\rangle$ and $|\psi^n_l\rangle$ are the $n_{th}$ standard right and left eigenvectors, respectively. The spectrum of eigenvalues can be separated into two parts with $Re[E]>0$ and $Re[E]<0$, which implies a line gap along $Re[E]$. $Re[E]$ represents the real part of the eigenvalue $E$. We suppose all the occupied states with $Re[E]<0$ ($Re[E^*]<0$) construct a matrix $U_r$ ($U_l$). On the other hand,  the matrix $V_r$ ($V_l$) constructed by the unoccupied states are shown in Fig.~\ref{f1}. Following the previous studies in Hermitian systems\cite{HOTI9,Qxy1,Qxy2,Qxy3,Qxy4}, we define the quadrupole moment in non-Hermitian systems as\cite{Qxy5}:
\begin{equation}
 Q_{xy}=\frac{1}{2\pi} {\rm Im \{log[det  } (U_l^\dagger \hat Q U_r)\sqrt{det(\hat Q^\dagger)}]\},
\label{QXY}
\end{equation}
with $\hat Q=exp[i2\pi \hat{x}\hat{y}/( N_x N_y)]$. $\hat{x}$ ($ N_x$) and $\hat{y}$ ($ N_y$) are the coordinate operator (sample size) along $x$ and $y$ directions, respectively. This definition makes use of the biorthogonal bases\cite{biorthogonal,NH6,NH7,dis6}, which is different from Hermitian cases.

Next, let us prove that the pseudoanti-Hermiticity\cite{NH10}[$H=-\eta H^\dagger\eta$, where the operator matrix $\eta$ is a real diagonal matrix with $\eta=\eta^{-1}$] guarantees the quantization of the quadrupole moment $ Q_{xy}$. The quantization of $ Q_{xy}$ is protected by the pseudoanti-Hermiticity instead of the chiral symmetry or particle-hole symmetry in Hermitian systems.
Similar to the previous study\cite{Qxy2}, one obtains
\begin{align}
\begin{split}
\det(U^\dagger_l\hat QU_r)&= \det[U_l^\dagger(\hat Q-\mathbb{I}+\mathbb{I})U_r]\\
&=\det[\mathbb{I}+U_l^\dagger(\hat Q-\mathbb{I})U_r]\\
&=\det[\mathbb{I}+(\hat Q-\mathbb{I})U_rU_l^\dagger],
\label{EQ1}
\end{split}
\end{align}
by using the Sylvester's determinant identity\cite{Qxy2} $\rm det(\mathbb{I}+AB)=\rm det(\mathbb{I}+BA)$.
 Biorthogonality-normalization relations indicate
\begin{equation}
\textbf{R}\textbf{L}^\dagger=
\left(
\begin{array}{cc}
U_r& V_r
\end{array}
\right)
\left(
\begin{array}{c}
U_l^\dagger \\
 V_l^\dagger\\
\end{array}
\right)=\mathbb{I}.
\end{equation}
Thus, one has $U_rU_l^\dagger=\mathbb{I}-V_rV_l^\dagger$. Noticing $V_l^\dagger V_r=\mathbb{I}$, one obtains:
\begin{align}
\begin{split}
\det(U^\dagger_l\hat QU_r)&=\det[\mathbb{I}+(\hat Q-\mathbb{I})(\mathbb{I}-V_rV_l^\dagger)]\\
&=\det[\hat Q-(\hat Q-\mathbb{I})V_rV_l^\dagger]\\
&=\det[\mathbb{I}+(\hat Q^\dagger-\mathbb{I})V_rV_l^\dagger]\det(\hat Q)\\
&=\det(V_l^\dagger \hat Q^\dagger V_r)\det(\hat Q).
\label{EQX}
\end{split}
\end{align}

After considering the pseudoanti-Hermiticity $\eta H \eta=-H^\dagger$ and Eq.~(\ref{EQ0}), one has
\begin{equation}
-H^\dagger[\eta |\psi_r^n\rangle]=E_n[\eta |\psi_r^n\rangle], ~-H[\eta |\psi_l^n\rangle]=E_n^*[\eta |\psi_l^n\rangle].
\label{EQ6}
\end{equation}
Comparing Eq.~(\ref{EQ6}) with Eq.~(\ref{EQ0}), one can find that $\eta |\psi_r^n\rangle$ and $\eta |\psi_l^n\rangle$ are the standard left and right eigenvectors with the corresponding eigenvalues $-E_n$ and $-E_n^*$, respectively. That is to say, if $E_n$ ($E_n^*$) is the eigenvalue of $H$ ($H^\dagger$), $\eta$ ensures that it has a corresponding partner $-E_n$ ($-E_n^*$), which is the eigenvalue of $H^\dagger$ ($H$).
The relationships between different eigenvalues and eigenvectors of $H$ and $H^\dagger$ are illustrated in Fig.~\ref{f1}. We have to point out that the Hermitian conjugation transformation used in this paper corresponds to the transformation between Hamiltonian $H$ and $H^\dagger$. Thus, $|\psi^n_r\rangle$ ($\eta |\psi_l^n\rangle$) will transform into $|\psi^n_l\rangle$ ($\eta |\psi_r^n\rangle$) under such a transformation, which is called Hermitian conjugation only for simplicity.

Provided that $Re[E_n]<0$, the following relations $|\psi_r^n\rangle\in U_r$, $|\psi_l^n\rangle\in U_l$, $\eta|\psi_r^n\rangle\in V_l$, and $\eta|\psi_l^n\rangle\in V_r$ can be obtained. These relations manifest that $\eta U_r= V_l$ as well as $\eta U_l=V_r$, and Eq.~(\ref{EQX}) can be rewritten as:
\begin{align}
\begin{split}
\det(U^\dagger_l\hat QU_r)&=\det(V_l^\dagger \hat Q^\dagger V_r)\det(\hat Q)\\
&=\det(U_r^\dagger\eta \hat Q^\dagger \eta U_l)\det(\hat Q).
\end{split}
\end{align}
Since $\eta$ is a real diagonal matrix and $\hat Q$ is a unitary-diagonal matrix, one obtaines $[\eta,\hat Q]=0$ and $\hat Q^\dagger \hat Q=\mathbb{I}$. Thus,
\begin{align}
\begin{split}
\det(U^\dagger_l\hat QU_r)\sqrt{\det(\hat Q^\dagger)}&=\det(U_r^\dagger \hat Q^\dagger U_l)\sqrt{\det(\hat Q)}\\
&=[\det(U^\dagger_l\hat QU_r)\sqrt{\det(\hat Q^\dagger)}~]^\dagger.
\label{EQ1}
\end{split}
\end{align}
Therefore, $\det(U^\dagger_l\hat QU_r)\sqrt{\det(\hat Q^\dagger)}$ is a real number, and $Q_{xy}$ should be quantized to $0$ or $0.5$ as expected.

\begin{figure}[t]
   \centering
    \includegraphics[width=0.43\textwidth]{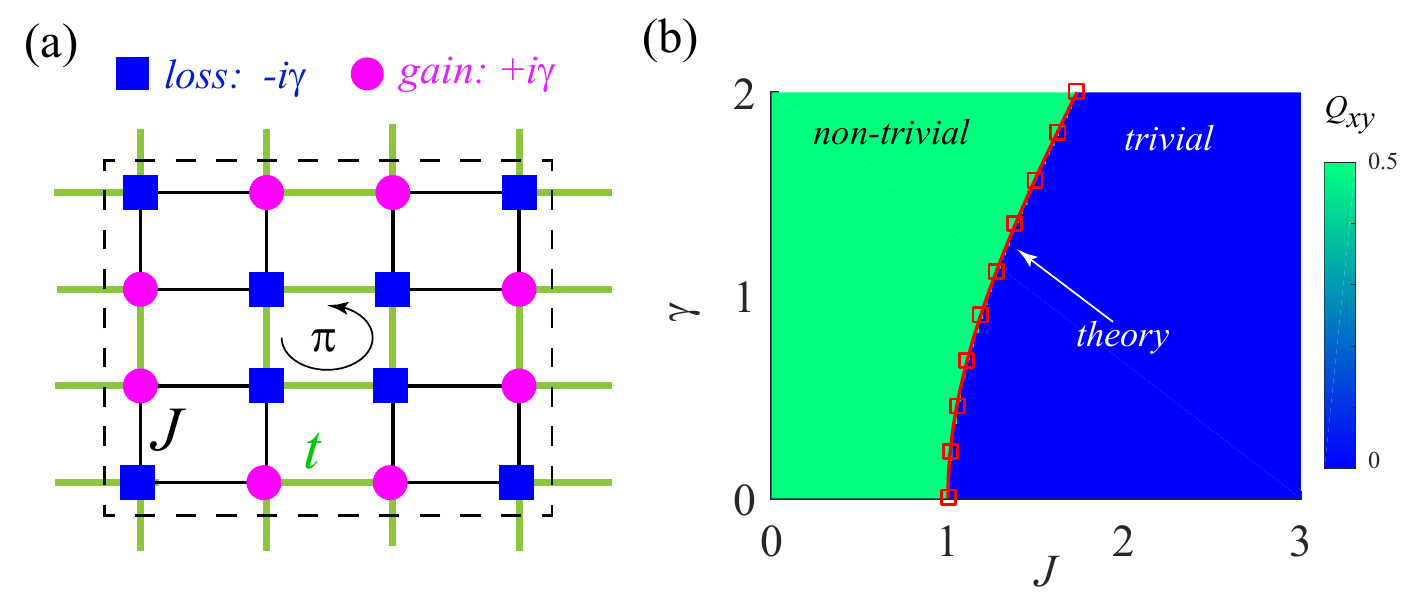}
    \caption{(Color online). (a) Schematic diagram of the tight-binding model. The blue square (pink circle) represents the sites with loss $-i\gamma$ (gain $i\gamma$). The green and black solid lines show the nearest hopping with strength $t$ and $J$, respectively. The dashed box is the primitive cell. $\pi$ flux is also considered. (b) The quadrupole moment $Q_{xy}$ versus $J$ and $\gamma$. The red solid line is obtained based on the theoretical analysis.  In all our calculation of $Q_{xy}$, the periodic boundary condition is adopted with sample size $N=40$.}
   \label{f2}
\end{figure}

\subsection{Model}

We consider a square-lattice model protected by the pseudoanti-Hermiticity as an example. The primitive cell contains $16$ sites, as shown in Fig.~\ref{f2}(a). It holds the pseudoanti-Hermiticity $\eta$, and the Hamiltonian reads\cite{NH10}:
\begin{align}
\begin{split}
\mathcal {H}=\sum_{n}i(\gamma+\varepsilon_n)\eta_n a_n^\dagger a_n+\sum_{\langle nm\rangle}t_{nm}e^{i\phi_{nm}}a^\dagger_na_m
\label{EQ9}
\end{split}
\end{align}
where $t_{nm}$ ( $J$ or $t$ ) represents the nearest-neighbor hopping strength between sites $n$ and $m$ with the lattice constant $a_0=1$. $J$ ($t$ ) corresponds to the black (green) solid lines in Fig.~\ref{f2}(a). $\eta_n$ represents the diagonal element $\eta(n,n)$ with the operator matrix of pseudoanti-Hermiticity $\eta=\sigma_z\sigma_z\sigma_z\sigma_z$. $a_n^\dagger$ ($a_n$) is the creation (annihilation) operator of site $n$. The magnetic flux\cite{phi} is determined by $\phi_{nm}=\frac{e}{\hbar}\int\textbf{A}\cdot\textbf{dl}$ with the vector potential $\textbf{A}=(-By,0,0)$, and $\textbf{dl}$ is the vector between $n$ and $m$ sites. We fix $\phi=a_0^2Be/\hbar=\pi$ and $t=1$ for simplicity. $\gamma$ denotes gain or loss strength marked in different colors, which is illustrated in Fig.~\ref{f2}(a).

 To maintain the pseudoanti-Hermiticity symmetry, the gain and loss disorder in one primitive cell is equivalent with $\varepsilon_n=\varepsilon$.  For different primitive cells, $\varepsilon$ is determined by Anderson disorder\cite{Anderson} with uniformly distributed $\varepsilon\in[-W/2,W/2]$, where $W$ is the disorder strength. Note that the system is set under half-filling conditions, i.e, the Fermi energy equals zero hereafter.

\section{numerical results}\label{S3}

This section illuminates that the real space topological invariant $Q_{xy}$ is a powerful tool to identify the higher-order topological nontrivial/trivial phases even when the non-Hermitian appears. We also investigate the disorder effect in the corresponding non-Hermitian systems.

\subsection{$Q_{xy}$ as an indicator of non-Hermitian HOTI}

Based on the quadrupole moment $Q_{xy}$ calculation under the periodic boundary condition, we first obtain a phase diagram of non-Hermitian Hamiltonian in Eq.~(\ref{EQ9}) with $W=0$. As plotted in Fig.~\ref{f2}(b), $Q_{xy}$ in the green region equals to $0.5$ (nontrivial) and changes into zero (trivial) in the blue area. The evolution of $Q_{xy}$ clearly shows the phase boundary shift between topological nontrivial and trivial phases by varying non-Hermitian strength $\gamma$. The red square solid line corresponds to the theoretically predicted phase boundary\cite{NH10} with $\gamma^2=2(J^2-1^2)$. Specifically, one can easily find that these two-phase boundaries obtained by different methods coincide perfectly.

\begin{figure}[b]
   \centering
    \includegraphics[width=0.40\textwidth]{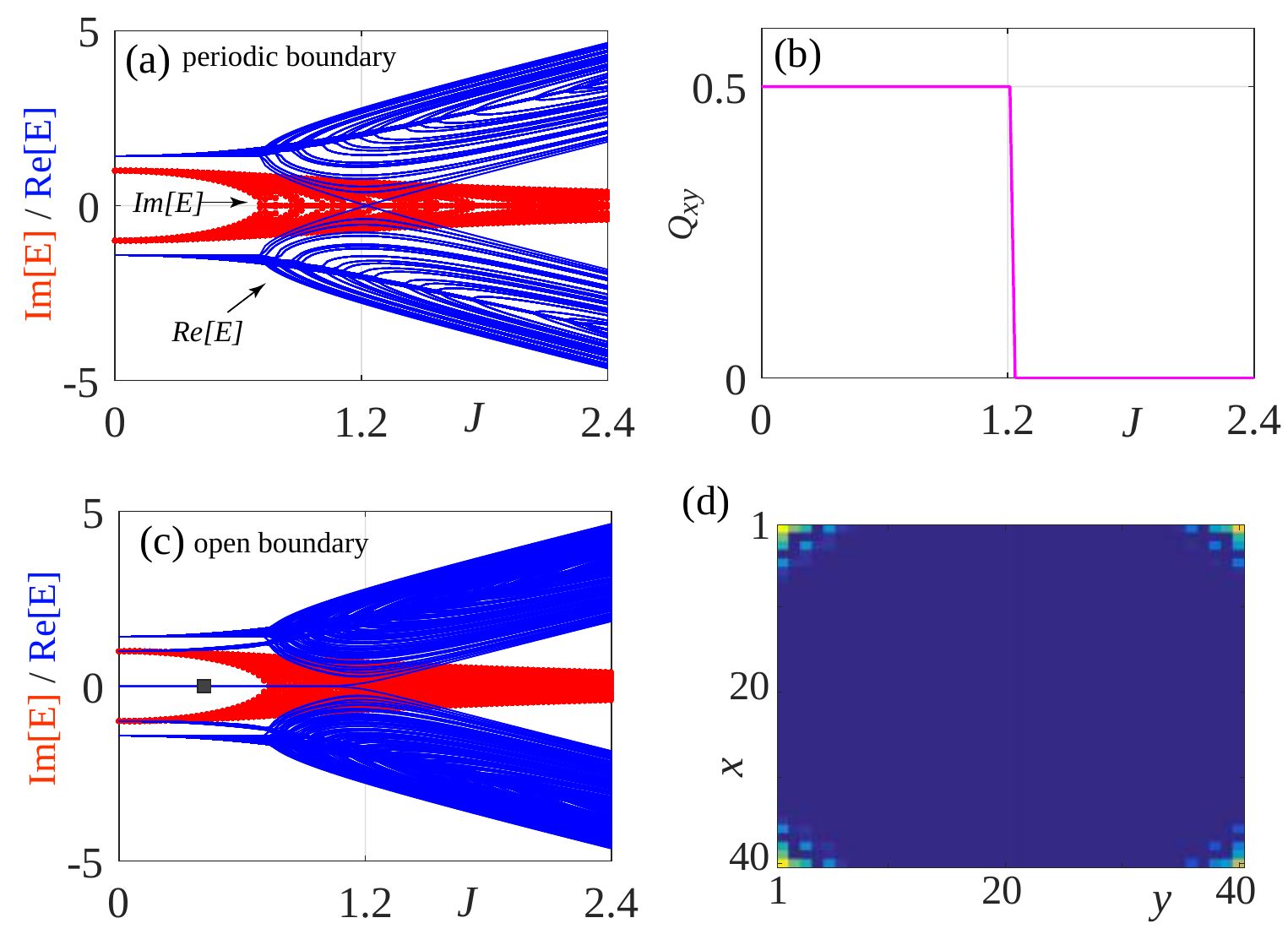}
    \caption{(Color online). (a) The plot of the real $Re[E]$ (image $Im[E]$) part of the eigenvalue $E$ versus $J$, marked in blue (red). (b) $Q_{xy}$ versus $J$. (c) The same with (a), except the open boundary is adopted. (d) A typical plot of the wavefunction, which is marked by black square shown in (c). We fix $\gamma=1$ in our calculation.}
   \label{f3}
\end{figure}

To better understand the nontrivial phase mentioned above, we analyze $Re[E]$ (blue) and $Im[E]$ (red) versus $J$ under the periodic boundary condition as well as the open one for $\gamma=1$. Theoretically, the bulk gap closes and reopens at $J=\sqrt{3/2}\approx 1.225 $ for $\gamma=1$. It consists with the plot of $Re[E]$ versus $J$ under the periodic boundary condition (see Fig.~\ref{f3}(a)). Meanwhile, $Q_{xy}$ versus $J$ curve (see Fig.~\ref{f3}(b)) shows that $Q_{xy}$ jumps from $0.5$ to $0$ at $J\approx\sqrt{3/2}$, which signals that the bulk gap closing induces a topological phase transition.
In order to clarify that $Q_{xy}=0.5$ determines the existence of non-Hermitian HOTI, we plot $Re[E]$ (blue) and $Im[E]$ (red) versus $J$ under the open boundary condition. Fig.~\ref{f3}(c) suggests that the zero-energy modes appear when $J<\sqrt{3/2}$, and the fourfold-degeneracy corner states emerge [see Fig.~\ref{f3}(d)].
The corner states' energy deviation from $Re[E]=0$ for $J\approx \sqrt{3/2}$ is due to the finite size effect.
 In short, the non-Hermitian HOTI exists when $J<\sqrt{3/2}$, and $Q_{xy}$ clearly distinguishes such a phase from the trivial one.

 To sum up, we show that the pseudoanti-Hermiticity symmetry protected quadrupole moment $Q_{xy}$ (defined as Eq.~(\ref{QXY})) can be utilized to distinguish non-Hermitian HOTI.

\subsection{Non-Hermitian HOTAI}

Subsequently, by considering disorder with pseudoanti-Hermiticity symmetry, we investigate the disorder-induced phase transition in non-Hermitian systems. Notably, we examine the existence of the non-Hermitian HOTAI.

\begin{figure}[t]
   \centering
    \includegraphics[width=0.49\textwidth]{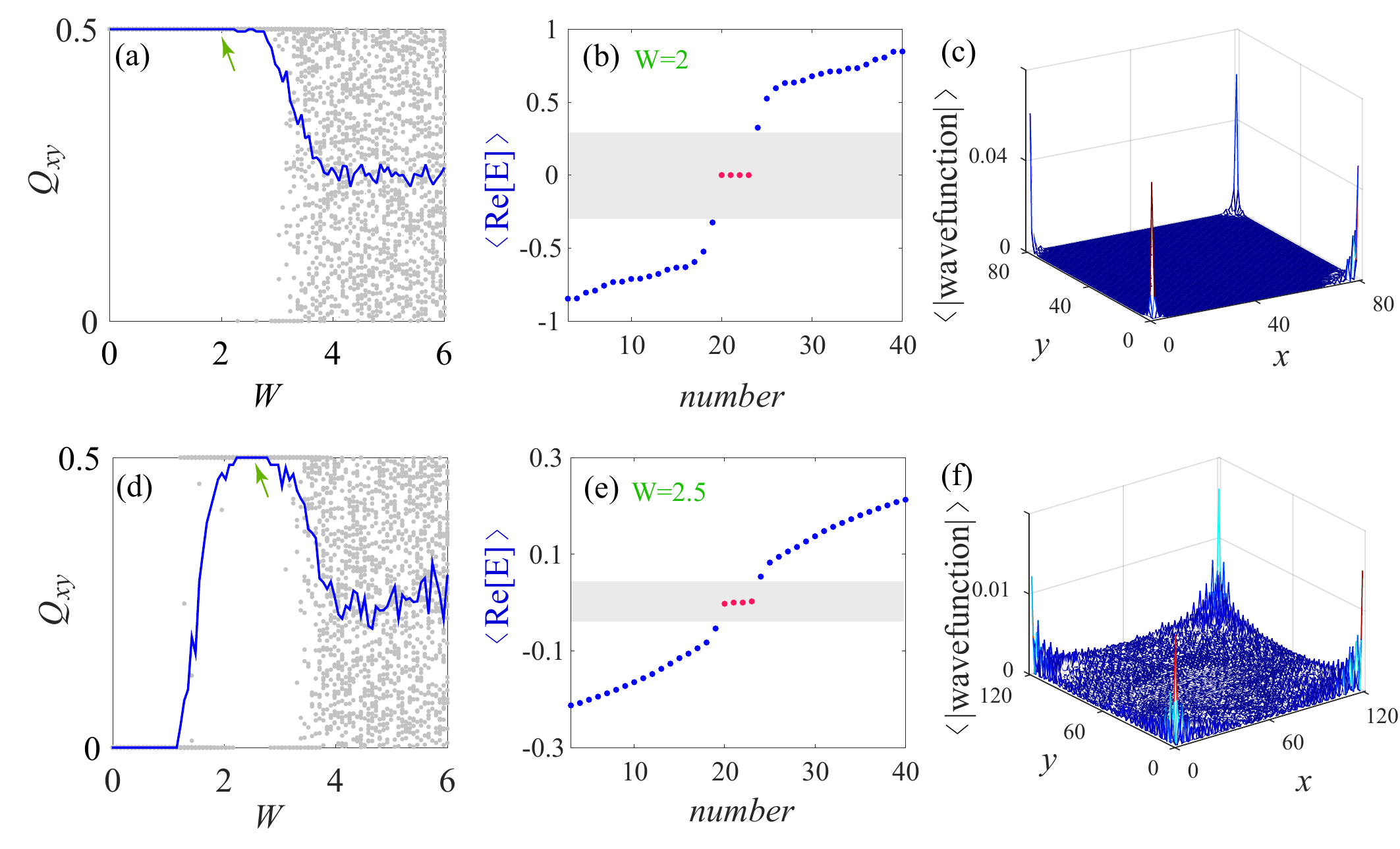}
    \caption{(Color online). (a) The evolution of quadrupole moment $Q_{xy}$ by increasing disorder strength $W$ with $J=0.5$. The gray dots are $Q_{xy}$ for each disordered sample, and the blue solid line is the average. (b) and (c) are the average energy $Re[E]$ and wavefunction [red dots marked in (b)] for $W=2$. (d)-(f) is similar to (a)-(c) except $J=1.26$ and $W=2.5$. $Q_{xy}$ is calculated under periodic boundary condition. Only several eigenvalues close to $Re[E]=0$ are plotted in (b) and (e). }
   \label{f4}
\end{figure}

The evolution of $Q_{xy}$ by increasing disorder strength $W$ with $J = 0.5$ is plotted in Fig.~\ref{f4}(a). Under the clean limit of $W=0$, the system is non-Hermitian HOTI with $Q_{xy}=0.5$. Unless otherwise specified, the disorder in numerical calculations holds the pseudoanti-Hermiticity symmetry $\eta$.
We find that the average quadrupole moment $Q_{xy}$ (solid blue line) is quantized to $0.5$ for $W<3$. Meanwhile, the zero-energy modes in the bulk gap [see Fig.~\ref{f4}(b)] as well as the corresponding wavefunction of corner states [see Fig.~\ref{f4}(c)] indicate that non-Hermitian HOTI is robust against disorder. Significantly, if the disorder-induced HOTAI emerges in such a system, it can also be identified by the quantization of $Q_{xy}$.

Next, we investigate the disorder-induced topological phase transition for $J=1.26$, as shown in Figs.~\ref{f4}(d)-(f). The sample is normal insulator (NI) with $Q_{xy}=0$ and without the zero-energy states under clean limit since $1.26>\sqrt{3/2}$.  With the increasing of disorder strength $W$, the average quadrupole moment $ Q_{xy}$ jumps from $0$ to $0.5$ [see Fig.~\ref{f4}(d)]. Further, the $ Q_{xy}=0.5$ plateau holds for disorder strength near $W\approx 2.5$.
Taking $W=2.5$ as an example, as shown in Fig.~\ref{f4}(e) and (f), we find the in-gap zero-energy states, as well as the corner states, exist. Different from those in Fig.~\ref{f4}(b) and (c), the four zero-energy modes in the bulk gap are no longer strictly degenerate, and the average wavefunction slightly penetrates into the bulk. These features should be related to the enhancement of coupling strength between four corner modes induced by disorder. However, the main property of the corner states in the bulk gap is still reserved. Therefore, it is reasonable to conclude that the region with $Q_{xy}=0.5$ in Fig.~\ref{f4}(d) belongs to the non-Hermitian HOTAI.

\begin{figure}[t]
   \centering
    \includegraphics[width=0.42\textwidth]{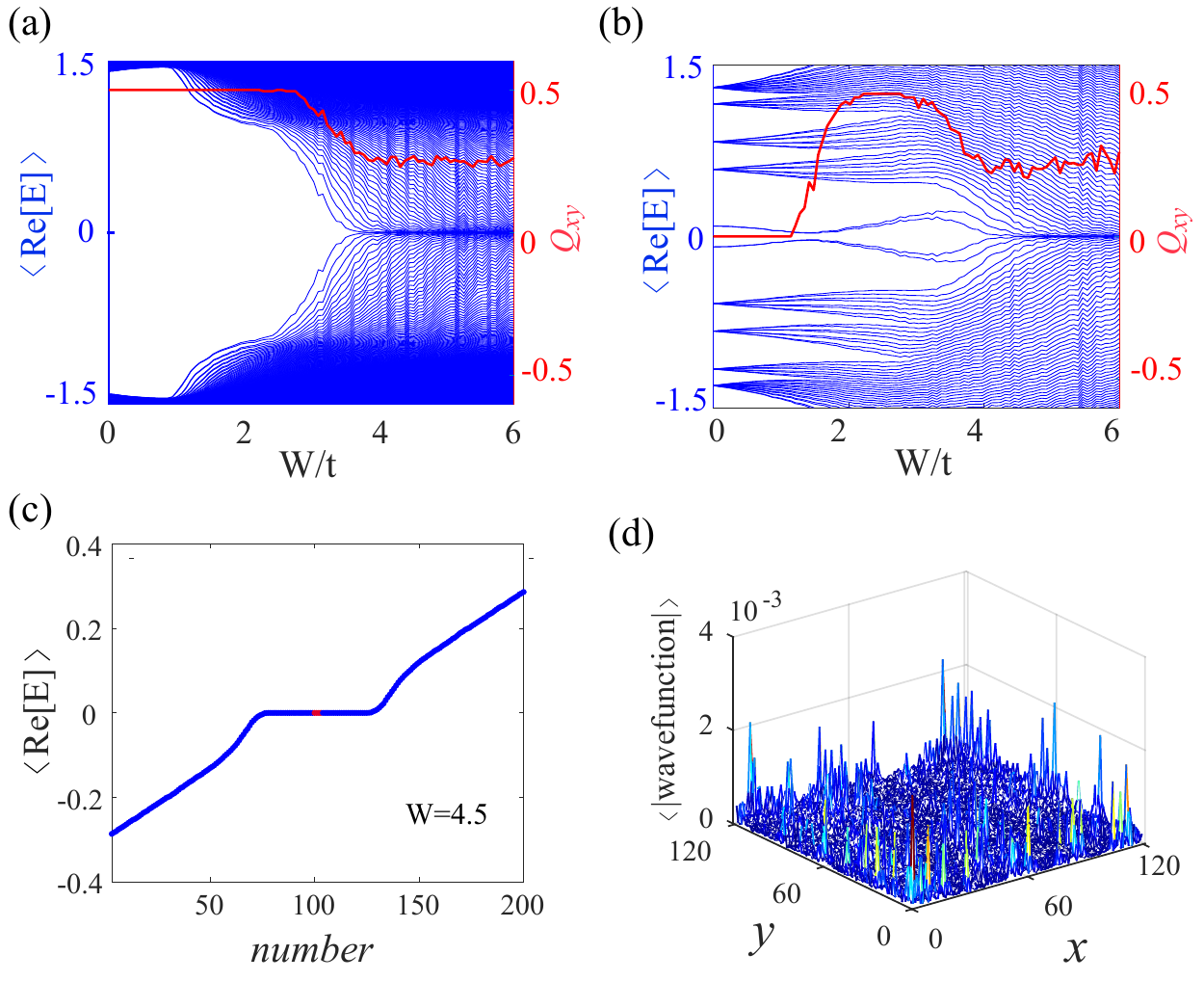}
    \caption{(Color online). (a) The evolution of $Re[E]$ with the increase of $W$. The red line shows $Q_{xy}$ for each disordered sample. Here, $J=0.5$. (b) is similar to (a) with $J=1.26$.
     (c) shows $Re[E]$ with disorder strength $W=4.5$. Only several eigenvalues close to $Re[E]=0$ are plotted.  (d) is the average wavefunction of the red dots in (c). (a) and (b) are calculated under periodic boundary condition with sample size $N=40$.  (c)-(d) are calculated under open boundary condition with $J=1.26$ and $N=120$. }
   \label{f5}
\end{figure}

 We also notice that $Q_{xy}$ cannot be quantized under strong disorder $W>3$ in both Fig.~\ref{f4}(a) and (d). The distribution of $Q_{xy}$ for each sample seems random, and the average quadrupole moment approximately equals 0.25. This behavior is also different from that in Hermitian HOTAI cases\cite{Qxy1,Qxy2}, where topological invariant $Q_{xy}$ decreases to zero under strong disorder. In order to explain the unquantized $Q_{xy}$ and further ensure the existence of the non-Hermitian HOTAI, we plot the evolution of $\langle Re[E]\rangle$ with the increase of $W$ under the periodic boundary condition [see Fig.~\ref{f5}(a) and (b)]. Here, $\langle Re[E]\rangle$ is the average of $Re[E]$.

  For $J=0.5$ (nontrivial case), the bulk gap gradually decreases and finally closes with the increasing disorder strength $W$ [see Fig.~\ref{f5}(a)]. $Q_{xy}$ is unquantized when $W>3$. Similarly, as shown in Fig.~\ref{f5}(b), we consider trivial case with $J=1.26$. The bulk gap closes and reopens with an increase of disorder strength. Meanwhile, $Q_{xy}$ jumps from $0$ to $0.5$, which strongly suggests the occurrence of the phase transition from NI to non-Hermitian HOTAI. Continuing to increase disorder strength for $W>3$, the bulk gap gradually reduces and closes. The quadrupole moment $Q_{xy}$ deviates from the quantized value accordingly.

In order to better understand the deviation of $Q_{xy}$ from the quantized value, we concentrate on the average $Re[E]$ as well as the average wavefunction with $W=4.5$. By setting $J=1.26$,
 the bulk gap disappears, and the average wavefunction extends into the bulk [see Figs.~\ref{f5}(c)-(d)]. Further, a series of $\langle Re[E]\rangle=0$ states appear, which are named as the gapless phase\cite{NH13}. The existence of states with $\langle Re[E]\rangle=0$ for the gapless phase is reasonable.  Taking
$
H=\left[
\begin{array}{cc}
i(\gamma+W) & t\\
t  & -i(\gamma+W)
\end{array}
\right]
$
as an intuitive example, its eigenvalues are $E=\pm\sqrt{t^2-(\gamma+W)^2}$. For fixed $t$ and $\gamma$, $\langle Re[E]\rangle$ tends to zero when $W$ is strong enough.
For such a case, we have to emphasize that the half-filling condition is not well-defined because of the existence of states with $Re[E]=0$ [see Fig.~\ref{f5}(c)]. However, the accuracy limitation in the numerical calculation slightly lifts these zero modes' degeneracy, which leads to an incorrect half-filling condition. Thus, the specific transformation between the occupied ($U_r,U_l$) and unoccupied states ($V_r,V_l$) cannot hold [see Fig.~\ref{f1}]. The unquantized $Q_{xy}$ can be considered as a signal for the existence of the gapless phase.


These findings indicate that the quantized $Q_{xy}$ requires not only symmetry but also the existence of bulk gap along $Re[E]$. When such a bulk gap disappears, the quantization of $Q_{xy}$ is not available. Significantly, the absence of bulk gap distinguishes the zero-energy modes in Fig.~\ref{f5}(c) from the non-Hermitian HOTAI shown in Fig.~\ref{f4}(e).
Moreover, the evolution of eigenvalues versus $W$ also strengthens the existence of non-Hermitian HOTAI shown in Figs.~\ref{f4}(d)-(f).

\section{summary and discussion}\label{S4}
In summary, based on the real space topological invariant [quadrupole moment in the frame of non-Hermitian Hamiltonian], we studied the disorder-induced HOTAI in non-Hermitian systems, which is protected by the pseudoanti-Hermiticity. First, we demonstrated that the quadrupole moment $Q_{xy}$ can be quantized to $0$ or $0.5$ by utilizing the biorthogonal bases. It is found that the quadrupole moment $Q_{xy}$ can be utilized to distinguish non-Hermitian HOTI ($Q_{xy}=0.5$) from the trivial ones ($Q_{xy}\neq0.5$). Then, with the help of $Q_{xy}$, we uncovered the disorder-induced phase transition in such a non-Hermitian system and demonstrated the existence of non-Hermitian HOTAI.

The real space topological invariant $Q_{xy}$ is a powerful tool to characterize non-Hermitian HOTI. In the appendix, we give another non-Hermitian HOTI without pseudoanti-Hermiticity symmetry and with NHSE to further declare the scope of application of $Q_{xy}$.
We show that the quantized $Q_{xy}$ is also available for samples with sublattice symmetry combined with a non-unitary transformation.
Besides, NHSE has no specific influence on the topological invariant $Q_{xy}$, except by considering the non-Bloch band theory\cite{NH6,NH7,NH8,NH9}.
Theoretically, the chiral symmetry and particle-hole symmetry in Altland-Zirnbauer classification for Hermitian cases should be replaced by specific symmetries in $38$-fold classification for non-Hermitian cases\cite{NH13}.
In addition to the pseudoanti-Hermiticity and sublattice symmetry reported in this paper, other symmetries in non-Hermitian systems may also be able to be utilized to ensure the quantization of $Q_{xy}$. Further, the line gap along $Im[E]$ will also influence the characterization of $Q_{xy}$. A much careful investigation can complete the understanding of topological features of non-Hermitian systems.

\section*{acknowledgement}
We are grateful to Yue-Ran Ding, Zibo Wang,  Ming Gong, and especially Qiang Wei for insightful discussion.
This work is financially supported by the National Basic Research Program of China No. 2019YFA0308403, National Natural Science Foundation of China No. 11822407, and a Project Funded by the Priority Academic Program Development of Jiangsu
Higher Education Institutions.


\setcounter{figure}{0}
\renewcommand{\thefigure}{A\arabic{figure}}
\appendix
\section{Influence of NHSE and the sublattice symmetry protected quadrupole moment in non-Hermitian systems}\label{appendix}

In the appendix, we investigate applicability of the quadrupole moment $Q_{xy}$ for other symmetries, and the NHSE is also considered. As an example, we focus on the sublattice symmetry protected non-Hermitian HOTI with NHSE. Although sublattice symmetry itself cannot ensure the quantization of $Q_{xy}$, we prove that there is another transformation, which connects $H$ to $H^{\dagger}$. The combination of these two transformations is similar to pseudoanti-Hermiticity, which connects $H$ to $-H^\dagger$. Therefore, the quantization of $Q_{xy}$ is still available even in the presence of disorder. Significantly, the quadrupole moment $Q_{xy}$ defined in real space is also applicable for systems with NHSE.

\subsection{ Quantization of $Q_{xy}$ for systems with sublattice symmetry and NHSE }

Following the previous study, the Hamiltonian for non-Hermitian HOTI with NHSE reads\cite{NH11}:
\begin{align}
\begin{split}
H\left(\textbf{k}\right)&=H_0+H\left(k_x\right)+H\left(k_y\right)\\
&=t\left(\tau_x\sigma_0+\tau_y\sigma_y\right)+i\gamma\left(\tau_y\sigma_x-\tau_y\sigma_z\right)\\
&+\lambda \cos\left(k_x\right)\tau_x\sigma_0-\lambda \sin\left(k_x\right)\tau_y\sigma_z\\
&+\lambda \cos\left(k_y\right)\tau_y\sigma_y+\lambda \sin\left(k_y\right)\tau_y\sigma_x,
\label{EQA1}
\end{split}
\end{align}
where $\tau_{x/y/z}$ and $\sigma_{x/y/z}$ are the Pauli matrices. $\tau_0$ and $\sigma_0$ are the identity matrices. $\gamma$ denotes the non-Hermitian strength. For simplicity, we fix $\gamma=0.4$. $H_0$ represents the intra primitive cell hopping matrix.
$H(k_x)$ and $H(k_y)$ show the hopping matrices between the nearest neighbor primitive cells along $x$ and $y$ directions, respectively.

 Eq.~(\ref{EQA1}) has the sublattice symmetry $SH\left(\textbf{k}\right)S=-H\left(\textbf{k}\right)$ with $S=\tau_z\sigma_0$. Sublattice symmetry operator $S$ connects $U_{l/r}$ to $V_{l/r}$ and vice versa.
 Eq.~(\ref{EQX}) implies that $S$ itself cannot ensure the quantization of $Q_{xy}$ since $U^\dagger_lU_l\neq \mathbb{I}$ in non-Hermitian systems\cite{Qxy2}. The quantization of $Q_{xy}$ demands another matrix $A$ which connects $H$ with $H^\dagger$. In addition, $A$ has to commutate with $\hat Q$ and satisfies $A^\dagger=A$. Moreover, the unbalanced hoppings along both $x$ and $y$ directions suggest the existence of NHSE.
In terms of two requirements above, we prove that there is a matrix, which transforms $H$ to $H^\dagger$ after eliminating the NHSE. Thus, the quantized $Q_{xy}$ is ensured.

In the following, we detail the derivation. The matrix forms of $H_0$, $H\left(k_x\right)$, and $H\left(k_y\right)$ are:
\begin{align}
\begin{split}
&H_0\equiv T_0=
\begin{bmatrix}
0&0&t-\gamma&-t+\gamma\\
0&0&t+\gamma&t+\gamma\\
t+\gamma&t-\gamma&0&0\\
-t-\gamma&t-\gamma&0&0
\end{bmatrix},\\
&H\left(k_x\right)= T_xe^{ik_x}+ T_x^\dagger e^{-ik_x},\\
&H\left(k_y\right)= T_ye^{ik_y}+ T_y^\dagger e^{-ik_y},\\
\end{split}
\end{align}
with
$
T_{x}=\begin{bmatrix}
0&0&\lambda&0\\
0&0&0&0\\
0&0&0&0\\
0&\lambda&0&0
\end{bmatrix}$
and
$T_{y}=\begin{bmatrix}
0&0&0&-\lambda\\
0&0&0&0\\
0&\lambda&0&0\\
0&0&0&0
\end{bmatrix}.
$
$T_{x/y}^\dagger$ is the hermitian conjugation of $T_{x/y}$. Due to the NHSE, the correct phase diagram is available by combining the non-Bloch band theory under periodic boundary condition. The replacement of momentum\cite{NH11} $k_{x/y}\rightarrow \tilde{k}_{x/y}+ik_0$ leads to
\begin{align}
\begin{split}
&H\left(\tilde{k}_x\right)=\beta T_xe^{i\tilde{k}_x}+\beta^{-1} T_x^\dagger e^{-i\tilde{k}_x},\\
&H\left(\tilde{k}_y\right)=\beta T_ye^{i\tilde{k}_y}+\beta^{-1} T_y^\dagger e^{-i\tilde{k}_y},\\
&H^\dagger\left(\tilde{k}_x\right)=\beta T_x^\dagger e^{-i\tilde{k}_x}+\beta^{-1} T_x e^{i\tilde{k}_x},\\
&H^\dagger\left(\tilde{k}_y\right)=\beta T_y^\dagger e^{-i\tilde{k}_y}+\beta^{-1} T_y e^{i\tilde{k}_y},
\end{split}
\end{align}
where $\beta=e^{-k_0}=\sqrt{|(t-\gamma)/(t+\gamma)|}$. Such replacement leaves $H_0$ unchanged.

Matrix $A$ should connect $H(\tilde{\textbf{k}})$ with $H^\dagger(\tilde{\textbf{k}})$.
Thus:
\begin{equation}
A
=
\begin{bmatrix}
\frac{t+\gamma}{t-\gamma}&0&0&0\\
0&\frac{t-\gamma}{t+\gamma}&0&0\\
0&0&1&0\\
0&0&0&1
\end{bmatrix}.
\end{equation}
It transforms $H_0$, $H\left(\tilde{k}_x\right)$ and $H\left(\tilde{k}_y\right)$ as follows:
\begin{align}
\begin{split}
&AH_0A^{-1}=H^\dagger_0,\\
&AH\left(\tilde{k}_x\right)A^{-1}=
a_1\beta T_xe^{i\tilde{k}_x}+a_2\beta^{-1} T_x^\dagger e^{-i\tilde{k}_x},\\
&AH\left(\tilde{k}_y\right)A^{-1}=
a_1\beta T_ye^{i\tilde{k}_y}+a_2\beta^{-1} T_y^\dagger e^{-i\tilde{k}_y},\\
\end{split}
\end{align}
with $a_1=(t+\gamma)/(t-\gamma)$, and $a_2=(t-\gamma)/(t+\gamma)$. Since $\beta=\sqrt{|(t-\gamma)/(t+\gamma)|}$, $a_1\beta$ and $a_2\beta^{-1}$ have two conditions to eliminate the absolute value sign of $\beta$. Two cases are analyzed as below:\\
 \textbf{case 1:} For $\frac{t-\gamma}{t+\gamma}>0$, $\beta=\sqrt{|\frac{t-\gamma}{t+\gamma}|}=\sqrt{\frac{t-\gamma}{t+\gamma}}$. The multiplication is available
\begin{align}
\begin{split}
a_1\beta=\beta^{-1},~
a_2\beta^{-1}=\beta.
\end{split}
\end{align}
It leads to
\begin{align}
\begin{split}
AH\left(\tilde{k}_x,\tilde{k}_y\right)A^{-1}=H^\dagger\left(\tilde{k}_x,\tilde{k}_y\right).
\end{split}
\end{align}
 \textbf{case 2:} For $\frac{t-\gamma}{t+\gamma}<0$, $\beta=\sqrt{|\frac{t-\gamma}{t+\gamma}|}=\sqrt{\frac{\gamma-t}{t+\gamma}}$. Thus,
\begin{align}
\begin{split}
a_1\beta=-\beta^{-1},~
a_2\beta^{-1}=-\beta.
\end{split}
\end{align}
One obtains
\begin{align}
\begin{split}
AH\left(\tilde{k}_x,\tilde{k}_y\right)A^{-1}=H^\dagger\left(\tilde{k}_x+\pi,\tilde{k}_y+\pi\right).
\end{split}
\end{align}

To make it more intuitive, let's analyze the transformation of $H\left(\tilde{k}_x,\tilde{k}_y\right)$ in the real space for a square sample with sample size $N\times N$. We suppose there is a diagonal matrix
\begin{equation}
A_R=diag(diag(A_1),diag(A_2),\cdots, diag(A_i),\cdots),
\end{equation}
 with $A_i=diag(diag(A_i))$. $diag$ represents the diagonal. $A_i$ is the construction block of $A_R$, which satisfies $A_i=\pm A$. To ensure $A_RHA_R^{-1}=H^\dagger$, the following relations should hold
\begin{align}
\begin{split}
&A_{i}T_{0}A_{i}^{-1}=T_{0}^\dagger;\\
&A_{i}\beta T_{x}A_{j}^{-1}=\beta^{-1} T_{x},~A_{i}\beta T_{y}A_{j}^{-1}=\beta^{-1} T_{y};\\
&A_{i}\beta^{-1} T_{x}^\dagger A_{j}^{-1}=\beta T_{x}^\dagger,~A_{i}\beta^{-1} T_{y}^\dagger A_{j}^{-1}=\beta T_{y}^\dagger;
\end{split}
\end{align}
where the corresponding hopping matrix between $i$ and $j$ primitive cells is clarified in the equation.
For all the $i\in[1,N^2]$,  $A_i=A$ when $\frac{t-\gamma}{t+\gamma}>0$. It ensures $A_RHA_R^{-1}=H^\dagger$.

When $\frac{t-\gamma}{t+\gamma}<0$,
we assume $A_{i\in odd}=-A_{j\in even}$ with $i,j\in[1,N]+kN$ and $k\in [0,N-1]$. For $i\in[1,N]+kN$ and $j\in[N+1,2N]+kN$ with $k\in [0,N-2]$, we require $A_{i}=-A_{j}$.
The above limits also preserve $A_RHA_R^{-1}=H^\dagger$.
Furthermore, we notice that $A_R$ is a diagonal matrix as expected for these two cases, and it commutates with $\hat Q$. Significantly, $A_R^\dagger=A_R$ is also satisfied.

Again, we take Eq.~(\ref{EQ0}) as an example and suppose $Re(E_n)<0$. $A_R$ satisfies $A_RHA_R^{-1}=H^\dagger$, which means:
\begin{align}
\begin{split}
&H^\dagger [A_R^{-1}|\psi^n_r\rangle]=E_n[A_R^{-1}|\psi^n_r\rangle],\\
&H [A_R|\psi^n_l\rangle]=E^*_n[A_R|\psi^n_l\rangle].
\end{split}
\end{align}
one obtains $A_R^{-1}|\psi^n_r\rangle\in U_l$ and $A_R|\psi^n_l\rangle\in U_r$. Noticing that the replacement $\textbf{k}\rightarrow \tilde{\textbf{k}}+ik_0$ maintains the sublattice symmetry $S$, the relationship $SHS=-H$ still holds with $S=S^{-1}$. Thus:
\begin{align}
\begin{split}
&-H [S|\psi^n_r\rangle]=E_n[S|\psi^n_r\rangle],\\
&-H^\dagger [S|\psi^n_l\rangle]=E^*_n[S|\psi^n_l\rangle].
\end{split}
\end{align}
It is obvious that $S|\psi^n_r\rangle\in V_r$ and $S|\psi^n_l\rangle\in V_l$.

\begin{figure}[t]
   \centering
    \includegraphics[width=0.47\textwidth]{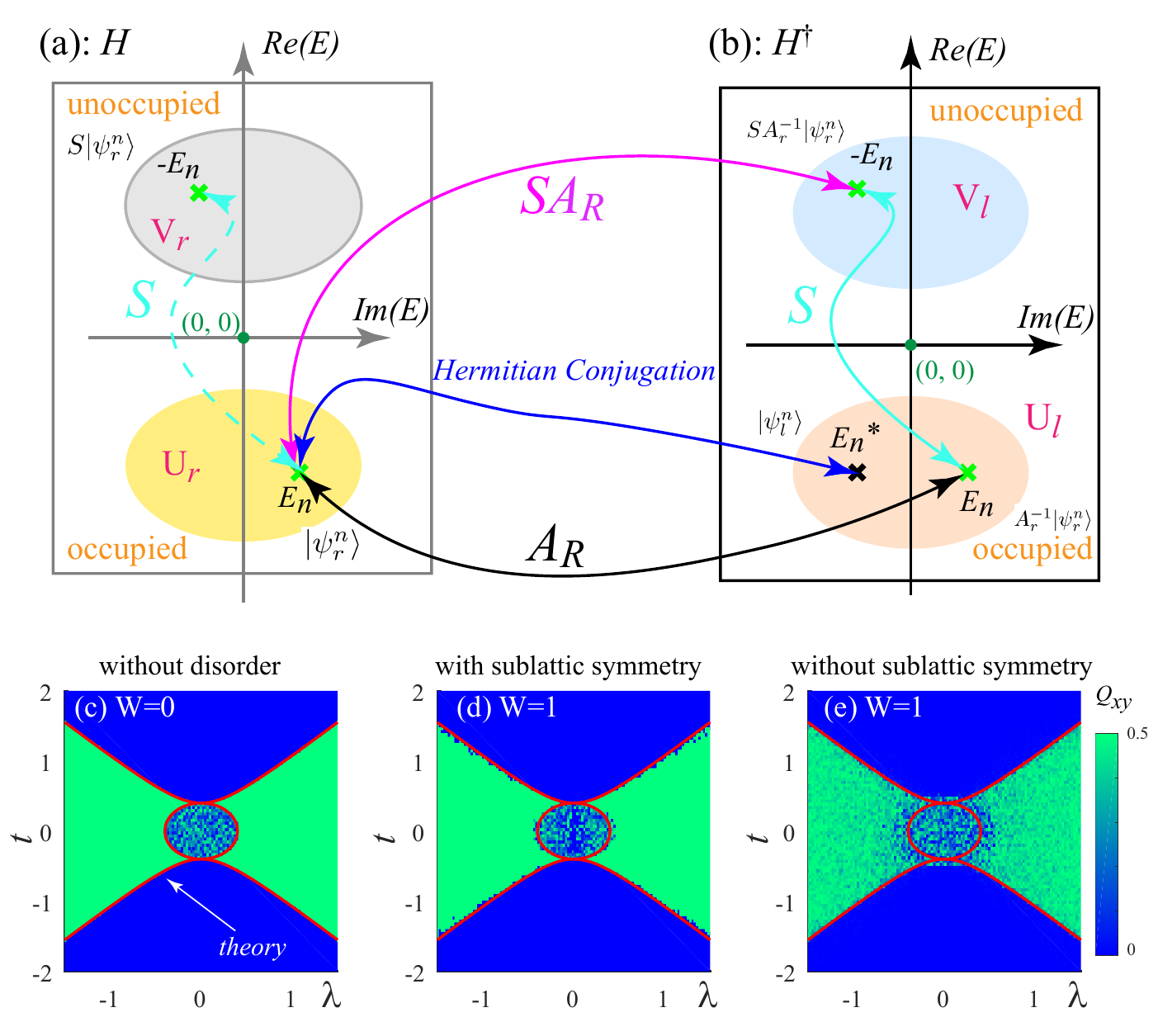}
    \caption{(Color online). (a) and (b) are similar to those in Fig.~\ref{f1}, except that the transformation operator is changed. $S$ corresponds to the sublattice symmetry. $A_R$ is a matrix defined in the appendix.
    (c)-(e) The phase diagram of non-Hermitian Hamiltonian Eq.~(\ref{EQA1}) with sublattice symmetry. The quadrupole moment $Q_{xy}$ versus $t$ and $\lambda$. The blue regions represent the topologically
trivial phase with $Q_{xy}=0$, while the green regions represent the non-Hermitian HOTI with $Q_{xy}=0.5$. The middle regions insides the ellipse represents the gapless states [$Q_{xy}$ lies between 0 and 0.5]. The phase boundaries\cite{NH11} (red line) are determined by $t^2=\gamma^2+\lambda^2$ and $t^2=\gamma^2-\lambda^2$ with $\gamma=0.4$. (c) with sublattice symmetry and without disorder $W=0$. (d) with sublattice symmetry and disorder strength $W=1$. (e) is similar to (d), except that the sublattice symmetry broken disorder is considered.}
   \label{f6}
\end{figure}

The corresponding relations between different states are shown in Fig.~\ref{f6}(a) and (b). For state $|\psi^n_r\rangle$ with eigenvalue $E_n$, $A_R$ will transform $|\psi^n_r\rangle\in U_r$ to $A_R^{-1}|\psi^n_r\rangle\in U_l$ with $E_n$ unchanged [the black solid line in Fig.~\ref{f6}(a) and (b)]. Then, the sublattice
 symmetry $S$ will transform $A_R^{-1}|\psi^n_r\rangle\in U_l$ to $SA_R^{-1}|\psi^n_r\rangle\in V_l$ with $E_n\rightarrow -E_n$ [the cyan solid line]. The combination of $S$ and  $A_R$ will connect $|\psi^n_r\rangle\in U_r$ to $SA_R^{-1}|\psi^n_r\rangle\in V_l$ [the pink solid line], which means $SA_R^{-1}$ connects $H$ to $-H^\dagger$.

We emphasize that the Hermitian-conjugation correlates $E_n$ (eigenvalue of $H$) with $E_n^*$ (eigenvalue of $H^\dagger$), which is different from the transformation $A_R$. $A_R$ correlates $E_n$ (eigenvalue of $H$) with $E_n$ (eigenvalue of $H^\dagger$). It means that the $n_{th}$ eigenvalue of $H^\dagger$ ($E_n$) should be a real number. Otherwise, $E_n$ must have a complex conjugate partner $E_n^*$, as shown in Fig.~\ref{f6}(b).

To sum up, we obtain the following transformations:
\begin{equation}
\left\{
\begin{array}{l}
SHS=-H \\
A_RHA_R^{-1}=H^\dagger
\end{array}\right.
\Rightarrow
\left\{
\begin{array}{l}
V_{l/r}=SU_{l/r} \\
U_l=A_R^{-1}U_r\\
U_r=A_RU_l
\end{array}\right..
\end{equation}
For non-Hermitian systems with sublattice symmetry, the quantization of $Q_{xy}$ is not guaranteed since $V_{l/r}=SU_{l/r}$ cannot ensure that $\det(U_r^\dagger \hat Q^\dagger U_l)\sqrt{\det(\hat Q)}$ is a real number. However, one can prove the quantization of $Q_{xy}$ is confirmed after combining $S$ and $A_R$. Following Eq.~(\ref{EQX}), we have:
\begin{align}
\begin{split}
\det(U^\dagger_l\hat QU_r)&=\det(V_l^\dagger\hat Q V_r)\det(\hat Q)\\
&=\det(U_l^\dagger S^\dagger \hat Q^\dagger S U_r)\det(\hat Q)\\
&=\det(U_l^\dagger \hat Q^\dagger U_r)\det(\hat Q)\\
&=\det(U_r^\dagger A_R^{-1\dagger} \hat Q^\dagger A_R U_l)\det(\hat Q)\\
&=\det(U_r^\dagger A_R^{-1} \hat Q^\dagger A_R U_l)\det(\hat Q)\\
&=\det(U_r^\dagger \hat Q^\dagger U_l)\det(\hat Q),
\end{split}
\end{align}
which means
\begin{equation}
\det(U^\dagger_l\hat QU_r)\sqrt{\det(\hat Q^\dagger)}=[\det(U_l^\dagger \hat Q U_r)\sqrt{\det(\hat Q^\dagger)}~]^\dagger.
\end{equation}
Thus, similar as Eq.~(\ref{EQ1}), the quantization of $Q_{xy}$ to $0$ or $0.5$ is ensured as expected.

\subsection{Numerical results}
We check that the quantized $Q_{xy}$ can be used to predict a correct phase diagram of such a non-Hermitian HOTI. The plots of $Q_{xy}$ versus $\gamma$ and $t$ are given in Fig.~\ref{f6}. To eliminate the influence of NHSE, the numerical calculation is based on $H({\tilde{k}_x,\tilde{k}_y})$ instead of $H(k_x,k_y)$. The periodic boundary condition is also adopted.

 When the disorder is absent, the quantized $Q_{xy}$ perfectly captures the existence of non-Hermitian HOTI. As shown in Fig.~\ref{f6}(c), the solid red line is the theoretically predicted phase boundary. The regions with $Q_{xy}=0.5$ fit the theoretical analysis quite well. We also notice that $ Q_{xy}$ shows different behaviors for different parameter regions. When $t^2>\gamma^2+\lambda^2$, one has $Q_{xy}=0$, which is a NI. However, $Q_{xy}$ is unquantized when $t^2<\gamma^2-\lambda^2$, and it seems to contradict with the symmetry-protected quantization of $Q_{xy}$. Actually, it corresponds to the gapless phase.

Next, we study the influence of disorder effect on the quantization of $Q_{xy}$. Two kinds of disorder, which preserves or breaks the sublattice symmetry, are considered. In Fig.~\ref{f6}(d), the disorder is added to the hopping terms along $y$ direction. Since the sublattice symmetry still holds, the quantized $Q_{xy}$ is still available. On the other hand, the quantized $Q_{xy}$ disappears [see Fig.~\ref{f6}(e)] when the on-site energy potential is adopted, which breaks the sublattice symmetry. These results are consistent with the previous analysis.


\begin{thebibliography}{999}
\bibitem{TI1} M. Z. Hasan, and C. L. Kane, Colloquium: topological insulators, Rev. Mod. Phys. \textbf{82}, 3045 (2010).
\bibitem{TI2} J. E. Moore, The birth of topological insulators, Nature \textbf{464}, 194 (2010).
\bibitem{TI3} X. L. Qi, and S. C. Zhang, Topological insulators and superconductors, Rev. Mod. Phys. \textbf{83}, 1057 (2011).
\bibitem{TI4} B. A. Bernevig, T. L. Hughes, and S. C. Zhang, Quantum Spin Hall Effect and Topological Phase Transition in HgTe Quantum Wells, Science \textbf{314}, 1757 (2006).
\bibitem{TI5} M. K$\ddot{o}$nig, S. Wiedmann, C. Br$\ddot{u}$ne, A. Roth, H. Buhmann, L. W. Molenkamp, X. L. Qi, and S. C. Zhang, Quantum Spin Hall Insulator State in HgTe Quantum Wells, Science \textbf{318}, 766 (2007).
\bibitem{TI6} L. Trifunovic, and P. W. Brouwer, Higher-Order Bulk-Boundary Correspondence for Topological Crystalline Phases, Phys. Rev. X \textbf{9}, 011012 (2019).
\bibitem{TI7} B. H. Yan, and C. Felser, Topological Materials: Weyl Semimetals, Annu. Rev. Condens. Matter Phys. \textbf{8}, 337 (2017).
\bibitem{TI8} Z. K. Liu, B. Zhou, Y. Zhang, Z. J. Wang, H. M. Weng, D. Prabhakaran, S. K. Mo, Z. X. Shen, Z. Fang, X. Dai, Z. Hussain, and Y. L. Chen, Discovery of a Three-Dimensional Topological Dirac Semimetal, Na3Bi, Science \textbf{343}, 864 (2014).
\bibitem{TI9} J. Alicea, New directions in the pursuit of Majorana fermions in solid state systems, Rep. Prog. Phys. \textbf{75}, 076501 (2012).

\bibitem{nature1} F. Tang, H. C. Po, A. Vishwanath, and X. G. Wan, Comprehensive search for topological materials using symmetry indicators, Nature \textbf{566}, 486 (2019).
\bibitem{nature2} T. T. Zhang, J. Jiang, Z. D. Song, H. Huang, Y. Q. He, Z. Fang, H. M. Weng, and C. Fang, Catalogue of topological electronic materials, Nature \textbf{566}, 475 (2019).
\bibitem{nature3} M. G. Vergniory, L. Elcoro, C. Felser, N. Regnault, B. A. Bernevig, and Z. J. Wang, The (High Quality) Topological Materials in the world, Nature \textbf{566}, 480 (2019).
\bibitem{Fu1} L. Fu, Topological Crystalline Insulators, Phys. Rev. Lett. \textbf{106}, 106802 (2011).
\bibitem{Fu2} Y. Ando, and L. Fu, Topological Crystalline Insulators and Topological Superconductors: From Concepts to Materials, Annu. Rev. Condens. Matter Phys. \textbf{6}, 361 (2015).


\bibitem{NH1}Z. P. Gong, Y. Ashida, K. Kawabata, K. Takasan, S. Higashikawa, and M. Ueda, Topological Phases of Non-Hermitian Systems, Phys. Rev. X \textbf{8}, 031079 (2018).
\bibitem{NH2}X. Y. Zhu, H. Q. Wang, S. K. Gupta, H. J. Zhang, B. Y. Xie, M. H. Lu, and Y. F. Chen Photonic non-Hermitian skin effect and non-Bloch bulk-boundary correspondence, Phys. Rev. Research \textbf{2}, 013280 (2020).
\bibitem{NH3}Q. B. Zeng, and Y. Xu, Winding numbers and generalized mobility edges in non-Hermitian systems, Phys. Rev. Research \textbf{2}, 033052 (2020).
\bibitem{NH4}K. Zhang, Z. S. Yang, and C. Fang, Correspondence between Winding Numbers and Skin Modes in Non-Hermitian Systems, Phys. Rev. Lett. \textbf{125}, 126402 (2020).
\bibitem{NH5}N. Okuma, K. Kawabata, K. Shiozaki, and M. Sato, Topological Origin of Non-Hermitian Skin Effects, Phys. Rev. Lett. \textbf{124}, 086801 (2020).
\bibitem{NH6}S. Y. Yao, and Z. Wang, Edge States and Topological Invariants of Non-Hermitian Systems, Phys Rev Lett. \textbf{121}, 086803 (2018).
\bibitem{NH7}S. Y. Yao, F. Song, and Z. Wang, Non-Hermitian Chern Bands, Phys. Rev. Lett. \textbf{121}, 136802 (2018).
\bibitem{NH8} K. Yokomizo, and S. Murakami, Non-Bloch Band Theory of Non-Hermitian Systems, Phys.Rev.Lett. \textbf{123}, 066404 (2019).
\bibitem{NH9} K. Yokomizo, and S. Murakami, Non-Bloch Band Theory and Bulk-Edge
Correspondence in Non-Hermitian Systems, arXiv:2009.04220 (2020).
\bibitem{NH10}  X.W. Luo, and C. W. Zhang, Higher-Order Topological Corner States Induced by Gain and Loss, Phys. Rev. Lett. \textbf{123}, 073601 (2019).
\bibitem{NH11}T. Liu, Y.R. Zhang, Q. Ai, Z. Gong, K. Kawabata, M. Ueda, and F. Nori, Second-order topological phases in non-Hermitian systems, Phys. Rev. Lett. \textbf{122}, 076801 (2019).

\bibitem{NH12}Y. Ashida, Z. P. Gong, and M. Ueda, Non-Hermitian Physics, arXiv: 2006.01837 (2020).
\bibitem{NH13} K. Kawabata, K. Shiozaki, M. Ueda, and M. Sato, Symmetry and Topology in Non-Hermitian Physics, Phys. Rev. X \textbf{9}, 041015 (2019).

\bibitem{NHSE1} K. Kawabata, M. Sato, and K. Shiozaki, Higher-order non-Hermitian skin effect, arXiv:2008.07237 (2020).
\bibitem{NHSE2} K. Zhang, Z. S. Yang, and C. Fang, Correspondence between winding numbers and skin modes in non-hermitian systems, arXiv:1910.01131 (2020).
\bibitem{NHSE4} Y. F. Yi, and Z. S. Yang, Non-Hermitian Skin Modes Induced by On-Site Dissipations and Chiral Tunneling Effect, Phys. Rev. Lett. \textbf{125}, 186802 (2020).
\bibitem{NHSE5} N. Okuma, and M. Sato, Quantum anomaly, non-Hermitian skin effects, and entanglement entropy in open systems, arXiv:2011.08175 (2020).
\bibitem{NHSE6} H. Jiang, L. J. Lang, C. Yang, S. L. Zhu, and S. Chen, Interplay of non-Hermitian skin effects and Anderson localization in nonreciprocal quasiperiodic lattices, Phys. Rev. B \textbf{100}, 054301 (2019).

\bibitem{dis1} A. F. Tzortzakakis, K. G. Makris, E. N. Economou, Non-Hermitian disorder in two-dimensional optical lattices, Phys. Rev. B \textbf{101}, 014202 (2020).
\bibitem{dis2} A. F. Tzortzakakis, K. G. Makris, S. Rotter, and E. N. Economou, Shape-preserving beam transmission through non-Hermitian disordered lattices, Phys. Rev. A \textbf{102}, 033504 (2020).
\bibitem{dis3} C. C. Wanjura, M. Brunelli, A. Nunnenkamp, Correspondence between non-Hermitian topology and directional amplification in the presence of disorder, arXiv:2010.14513 (2020).
\bibitem{dis4} A. F. Tzortzakakis, K. G. Makris, A. Szameit, and E. N. Economou, Non-Hermitian lattices with binary-disorder, arXiv:2007.08825 (2020)
\bibitem{dis5} L. Z. Tang, L. F. Zhang, G. Q. Zhang, and D. W. Zhang, Topological Anderson insulators
    in two-dimensional non-Hermitian disordered systems, Phys. Rev. A \textbf{101}, 063612 (2020).
\bibitem{dis6}H. F. Liu, Z. X. Su, Z-Q. Zhang, and H. Jiang, Topological Anderson insulator in two-dimensional non-Hermitian systems, Chinese Phys. B \textbf{29}, 050502 (2020).

\bibitem{NHD1}D. W. Zhang, L. Z. Tang, L. J. Lang, H. Yan, and S. L. Zhu, Non-Hermitian Topological Anderson Insulators, Sci. China-Phys. Mech. Astron. \textbf{63}, 267062 (2020).
\bibitem{NHD2} X. W. Luo, and C. W. Zhang, Non-Hermitian Disorder-induced Topological insulators, arXiv:1912.10652 (2019).
\bibitem{IPR1}C. Wang and X. R. Wang, Level statistics of extended states in random non-Hermitian Hamiltonians, Phys. Rev. B \textbf{101}, 165114 (2020).
\bibitem{XunlongLuo} X. L. Luo,1, T. Ohtsuki, and R. Shindou, Universality classes of the Anderson Transitions Driven by non-Hermitian Disorder, arXiv:2011.07528.
  
  
    
\bibitem{NGM2} J. T. Song, and E. Prodan, AIII and BDI topological systems at strong disorder, Phys. Rev. B \textbf{89}, 224203 (2014).
\bibitem{NGM3} Z. Q. Zhang, C. Z. Chen, Y. J. Wu, H. Jiang, J. W. Liu, Q. F. Sun, and X. C. Xie, Chiral Interface States and Related Quantized Transport in Disordered Chern Insulators, Phys. Rev. B \textbf{103}, 075434 (2021).
\bibitem{NGM4} Z. X. Su, Y. Z. Kang,  B. F. Zhang, Z. Q. Zhang and H. Jiang, Disorder induced phase transition in magnetic higher-order topological insulator: A machine learning study, Chin. Phys. B \textbf{28}, 117301 (2019).
\bibitem{NGM5} E. Prodan, Disordered topological insulators: a non-commutative geometry perspective, Phys. A-Math. Theor. \textbf{44}, 113001 (2011).
\bibitem{NGM6} Y. F. Zhang, Y. Y. Yang, Y. Ju, L. Sheng, R. Shen, D. N. Sheng, and D. Y. Xing, Coupling-matrix approach to the Chern number calculation in disordered systems, Chin. Phys. B \textbf{22} 117312 (2013).

\bibitem{HOTI1} J. Langbehn, Y. Peng, L. Trifunovic, F. von Oppen, and P. W. Brouwer, Reflection-Symmetric Second-Order Topological Insulators and Superconductors, Phys. Rev. Lett. \textbf{119}, 246401 (2017).
\bibitem{HOTI2} W. A. Benalcazar, B. A. Bernevig, and T. L. Hughes, Quantized electric multipole insulators, Science \textbf{357}, 61 (2017).
\bibitem{HOTI3} C. Shang, X. N. Zang, W. L. Gao, U. Schwingenschl$\ddot{o}$gl, and A. Manchon, Second-order topological insulator and fragile topology in topological circuitry
simulation, arXiv:2009.09167 (2020).
\bibitem{HOTI4} W. A. Benalcazar, B. A. Bernevig, and T. L. Hughes, Electric Multipole Moments, Topological Multipole Moment Pumping, and Chiral Hinge States in Crystalline Insulators, Phys. Rev. B \textbf{96}, 245115 (2017).
\bibitem{HOTI5} F. Schindler, A. M. Cook, M. G. Vergniory, Z. J. Wang, S. S. P. Parkin, B. A. Bernevig, and T. Neupert, Higher-order topological insulators, Science Advances. \textbf{4} (2018).
\bibitem{HOTI6} M. Geier, L. Trifunovic, M. Hoskam, and P. W. Brouwer,  Second-order topological insulators and superconductors with an order-two crystalline symmetry, Phys. Rev. B \textbf{97}, 205135 (2018).
\bibitem{HOTI7}M. Ezawa, Higher-Order Topological Insulators and Semimetals on the Breathing Kagome and Pyrochlore Lattices, Phys. Rev. Lett. \textbf{120}, 026801 (2018).
\bibitem{HOTI8}C.-H. Hsu, P. Stano, J. Klinovaja, and D. Loss, Majorana Kramers Pairs in Higher-Order Topological Insulators, Phys. Rev. Lett. \textbf{121}, 196801 (2018).
\bibitem{HOTI9}B. Kang, K. Shiozaki, and G. Y. Cho, Many-body order parameters for multipoles in solids, Phys. Rev. B \textbf{100}, 245134 (2019).

\bibitem{Qxy1} Y. B Yang, K. Li, L. M. Duan, and Y. Xu, Higher-order Topological Anderson Insulators, Phys. Rev. B \textbf{103}, 085408 (2021).
\bibitem{Qxy2}  C. A. Li , B Fu, Z. A. Hu, J. Li , and S. Q. Shen, Topological Phase Transitions in Disordered Electric Quadrupole Insulators, Phys. Rev. Lett. \textbf{125}, 166801 (2020).
\bibitem{Qxy3}W. A. Wheeler, L. K. Wagner, and T. L. Hughes, Manybody electric multipole operators in extended systems, Phys. Rev. B \textbf{100}, 245135 (2019).
\bibitem{Qxy4}B. Roy, Antiunitary symmetry protected higher-order topological phases, Phys. Rev. Research \textbf{1}, 032048 (2019).
\bibitem{Qxy5}H. Wu, B. Q. Wang, and J. H. An, Floquet second-order topological insulators in non-Hermitian systems, Phys. Rev. B \textbf{103}, L041115 (2021).

\bibitem{NHHO1}Z. Zhang, M. R. L$\acute{o}$pez, Y. Cheng, X. Liu, J. Christensen, Non-Hermitian Sonic Second-Order Topological Insulator, Phys. Rev. Lett. \textbf{122}, 195501 (2019).
\bibitem{NHHO2} E. Edvardsson, F. K. Kunst, E. J. Bergholtz, Non-Hermitian extensions of higher-order topological phases and their biorthogonal bulk-boundary correspondence,Phys. Rev. B \textbf{99}, 081302 (2019).

\bibitem{biorthogonal} D. C. Brody, Biorthogonal quantum mechanics, J. Phys.A: Math. Theor. \textbf{47}, 035305 (2013).

\bibitem{phi} W. Long, Q. F. Sun, and J. Wang, Disorder-Induced Enhancement of Transport through Graphene p-n Junctions, Phys. Rev. Lett. \textbf{101}, 166806 (2008).
\bibitem{Anderson} P. W. Anderson, Absence of Diffusion in Certain Random Lattices, Phys. Rev. \textbf{109}, 1492-1505 (1958).

\end{thebibliography}
\end{document}